\documentclass[11pt,a4wide]{article}

\usepackage{amsfonts}
\usepackage{amsbsy}
\usepackage{epsfig}
\usepackage{latexsym}
\input amssym.def
\input amssym.tex

\advance \topmargin by -\headheight
\advance \topmargin by -\headsep
\evensidemargin \oddsidemargin
\advance\hoffset by -3mm  
\advance\voffset by  8mm  
\def\bbbc{{\mathchoice {\setbox0=\hbox{$\displaystyle\rm C$}\hbox{\hbox
to0pt{\kern0.4\wd0\vrule height0.9\ht0\hss}\box0}}
{\setbox0=\hbox{$\textstyle\rm C$}\hbox{\hbox
to0pt{\kern0.4\wd0\vrule height0.9\ht0\hss}\box0}}
{\setbox0=\hbox{$\scriptstyle\rm C$}\hbox{\hbox
to0pt{\kern0.4\wd0\vrule height0.9\ht0\hss}\box0}}
{\setbox0=\hbox{$\scriptscriptstyle\rm C$}\hbox{\hbox
to0pt{\kern0.4\wd0\vrule height0.9\ht0\hss}\box0}}}}

\makeatletter
\@addtoreset{equation}{section}
\makeatother

\begin{document}

\hfuzz=100pt \title{{\Large \bf{Spherical Spacelike Geometries in Static Spherically Symmetric Spacetimes: Generalized Painlev\`{e}-Gullstrand Coordinates, Foliation, and Embedding
 }}}
\author{\\M M Akbar\footnote{E-mail: akbar@utdallas.edu} \\
Department of Mathematical Sciences\\ University of Texas at Dallas\\ Richardson, TX 75080, USA}
 \date{\today} \maketitle
\begin{abstract}
It is well known that static spherically symmetric spacetimes can admit foliations by flat spacelike hypersurfaces, which are best described in terms of the Painlev\`{e}--Gullstrand coordinates. The uniqueness and existence of such foliations were addressed earlier. In this paper, we prove, purely geometrically, that any possible foliation of a static spherically symmetric spacetime by an arbitrary codimension-one spherical spacelike geometry, up to time translation and rotation, is unique, and we find the algebraic condition under which it exists. This leads us to what can be considered as the most natural generalization of the Painlev\`{e}--Gullstrand coordinate system for static spherically symmetric metrics, which, in turn, makes it easy to derive generic conclusions on foliation and to study specific cases as well as to easily reproduce previously obtained generalizations as special cases.
In particular, we note that the existence of foliation by flat hypersurfaces guarantees the existence of foliation by hypersurfaces whose Ricci curvature tensor is everywhere non-positive (constant negative curvature is a special case). The study of uniqueness and the existence concurrently solves the question of embeddability of a spherical spacelike geometry in one-dimensional higher static spherically symmetric spacetimes, and this produces known and new results geometrically, without having to go through the momentum and Hamiltonian constraints.
\end{abstract}
\section{Introduction}
As the first nontrivial solution of the vacuum Einstein equations, the Schwarzschild metric
\begin{equation}
ds^2= -\left(1-\frac{2m}{r}\right)dt^2 + \frac{dr^2}{1-\frac{2m}{r}}+ r^2\left(d\theta^2 + \sin^2{\theta} d\phi^2\right), \label{Schw}
\end{equation}
captures all of our immediate intuitive expectations of the gravitational field outside a spherical symmetric star -- its components are independent of the time coordinate $t$ and depend only on the radial coordinate $r$ in spherical coordinates. This metric is singular at $r=2m$ and, thus, it is limited to only $r \in (2m, \infty)$. Moving to suitable coordinate systems, however, one finds that this is, in fact, a coordinate singularity, and one can include the $0<r \le 2m$ region of the spacetime, i.e., we can ``extend" (\ref{Schw}), leaving $r=0$ as the only true singularity of the spacetime with $r=2m$ a ``horizon." In most expositions of relativity, one uses the Eddington--Finkelstein or Kruskal--Szekeres coordinate systems to show this, in both of which one necessarily mixes the old radial and time coordinates. However, an even older coordinate system, which was introduced by Painlev\`{e} and Gullstrand independently for advancing different arguments, achieves this \cite{PaulPainleve1921, Gullstrand1922}. In this too, one mixes the original time and radial coordinates. Although originally written for the Schwarzschild metric, this can easily be extended to any spherically symmetric static spacetime in arbitrary dimensions. In recent years, this coordinate system has resurfaced in connection with both classical and quantum aspects of gravity \cite{Kraus-Wilczek, Parikh-Wilczek, Hawk-Hun}. Quite independently, it has also appeared in various gravity analog systems in condensed matter physics (see for example, \cite{Volovik}, and references therein).

An additional insight that comes with Painlev\`{e}--Gullstrand coordinates -- something that the other coordinates mentioned above do not reveal -- is the quite generic existence of flat hypersurfaces in spherically symmetric static spacetimes (see below). For the Schwarzschild, these flat hypersurfaces provide a foliation. For a general static spherically symmetric spacetime, such a foliation is unique up to time translation (and rotational symmetry) \cite{Beig-Siddiqui}. However, unlike the Schwarzschild, in a general static spherically symmetric spacetime such a foliation exists only under certain algebraic conditions of the original metric components \cite{Soo}. The existence of flat foliations in spherical gravity can also be seen from the Hamiltonian formulation of general relativity \cite{Guven:1999hc,Husain:2001wk}.

In this paper, we first prove that spherically symmetric hypersurfaces in static spherically symmetric spacetimes are unique up to time translation and rotational symmetry. Our generalized Painlev\`{e}--Gullstrand coordinates then follow from choosing coordinates adapted to these hypersurfaces, in the same way the original Painlev\`{e}--Gullstrand coordinates correspond to the zero-curvature hypersurfaces. The algebraic conditions under which these hypersurfaces provide a foliation follows from the latter.
The embeddability of an arbitrary codimension-one spherical geometry, previously studied for specific spacetimes via Hamiltonian and momentum constraints of spherical general relativity, follows purely geometrically from the existence condition of foliation (with a simple interpretational change in the existence condition) from which many new examples as well as known ones can quickly be worked out.
\section{Flat Hypersurfaces and Painlev\`{e}--Gullstrand Coordinates}
In the (original) Painlev\`{e}--Gullstrand coordinates for the Schwarzschild metric (\ref{Schw}), the new time coordinate is defined in terms of the original time and radial coordinates, as follows:
\begin{equation}
\overline{t}=t+ 4m \left[ \sqrt{\frac{r}{2m}}+\frac{1}{2}\ln \left(\frac{\sqrt{\frac{r}{2m}}-1}{\sqrt{\frac{r}{2m}}+1}\right)\,\right],
\end{equation}
which turns (\ref{Schw}) into
\begin{equation}
ds^2= -\left(1-\frac{2m}{r}\right)d\overline{t}^2 + 2\sqrt{\frac{2m}{r}} dr d\overline{t}+ dr^2 + r^2(d\theta^2 + \sin^2{\theta} d\phi^2). \label{SchPG}
\end{equation}
It is easy to see that (\ref{SchPG}) is regular across $r=2m$ and that a $\overline{t}=\mathrm{const.}$ hypersurface of (\ref{SchPG}) is a three-dimensional flat metric expressed in polar coordinates. These hypersurfaces run across the horizon, $r=2m$, and they have only $r=0$ as a coordinate singularity. Note that, because $r=0$ is a genuine singularity, the $r=0$ point is {\it{a priori}} absent from the spacetime and cannot be included, unlike the apparent $r=2m$ singularity. Hence, $\overline{t}=\mathrm{const.}$ hypersurfaces are all, in fact, $\mathbb{R}^3-\{0\}$, i.e., Euclidean three-spaces with a point removed. This does not show up in the Painlev\`{e}--Gullstrand form, (\ref{SchPG}), where $r=0$ has the appearance of a polar coordinate singularity, which probably explains why this was not given much attention in the literature.

For any static spherically symmetric metric in arbitrary $(d+1)$ dimensions,
\begin{equation}
ds^2= -f(r) dt^2+ \frac{1}{g(r)} dr^2 +r^2 d\Omega_{d-1}^2 \label{GenStat},
\end{equation}
one can likewise take
\begin{equation}
\overline{t}=t+\int\sqrt{\frac{1-g}{fg}}\,dr,
\end{equation}
equivalently,
\begin{equation}
d\overline{t}=dt+\sqrt{\frac{1-g}{fg}}\,dr, \label{diffrelation}
\end{equation}
and write (\ref{GenStat}) as
\begin{equation}
ds^2= -f(r)d\overline{t}^2 + 2\sqrt{\frac{f(1-g)}{g}} dr d\overline{t}+ dr^2 + r^2d\Omega_{d-1}^2, \label{GenStatPG}
\end{equation}
in which $\overline{t}=\mathrm{const.}$ hypersurfaces are flat.
\subsubsection*{Existence and Uniqueness}
However, the $\overline{t}=\mathrm{const.}$ flat hypersurfaces in (\ref{GenStatPG}) exist provided the algebraic condition relating the metric components of the original (\ref{GenStat}) is satisfied:
\begin{equation}
\frac{f(1-g)}{g} \ge 0.\label{algcondflat}
 \end{equation}
Assuming $\frac{f}{g}>0$, this condition requires $g < 1$, and both of these, for example, are fulfilled by the Schwarzschild metric (\ref{Schw}).  Are these hypersurfaces in static spherically symmetric spacetimes unique? This was answered affirmatively in \cite{Beig-Siddiqui}. The coordinates adapted to their uniqueness proof are nothing but the Painlev\`{e}--Gullstrand (\ref{GenStatPG}).  The algebraic condition for existence, complementing the uniqueness proof in \cite{Beig-Siddiqui}, was discussed in \cite{Soo}. As an example, \cite{Soo} noted that (their equivalent of) condition (\ref{algcondflat}) fails to hold in the Reissner--Nordstr{\"{o}}m and Schwarzschild--anti de Sitter metrics at small and large values of $r$, respectively. Since $\frac{f}{g}>0$, this means $g < 1$ does not hold for these metrics.
\section{Uniqueness of Constant-curvature Hypersurfaces}
In connection with the above work, \cite{Soo} considered how an arbitrary spherically symmetric metric transforms under a general local Lorentz boost.\footnote{Initially, \cite{Soo} considered the
time-dependent version of {(\ref{GenStat})} (i.e.\ $f$ and $g$ are
functions of $t$ and $r$), starting with a Painlev\`e--Gullstrand time
(in our notation)
\begin{equation}
d\overline{t}=dt+ \beta (t, r) dr.
\end{equation}
However, the vanishing of the exterior derivative of this equation
requires that \mbox{$\frac{\partial \beta }{\partial t} =0$}, forcing
$\beta (t, r)=\beta (r)$, which in turn, requires (in our notation)
$g$ to be a function of $r$ alone. Thus, one is left with only the
static case. This was addressed in the final arXiv version of
the paper in a footnote. The
generalized Painlev\`e--Gullstrand form obtained by them continues to hold provided its
components are read as functions of $r$ only; most of their
examples and discussions were on the static case and hence do not change. This is
what is relevant for us too. In any case, for the
time-dependent case, one can choose a different
(non-Painlev\`e--Gullstrand) time,
\begin{equation}
dt_{N}=\alpha (t, r) dt + \beta (t, r) dr,
\end{equation}
provided $\frac{\partial \alpha }{\partial r} =\frac{\partial \beta }{
\partial t}$. This is equivalent to the integrating factor
mentioned in the footnote in the final arXiv version. We thank Chopin Soo for
referring us to it.} This essentially mixes $dt$ with $dr$ and changes the warping function of the spherical part ($=r$ in our notation) to a more general function. The transformed metric contains the Eddington--Finkelstein as a special case, but generally it can be described as a generalization of the Painlev\`{e}--Gullstrand coordinates for static spherically symmetric spacetimes, from which, for example, the existence condition for the flat slices follows. They, in particular, examined the existence conditions of constant-curvature hypersurfaces showing that  these can potentially exist when flat slices do not (and, thus, they can avoid the problems posed by flat foliation in the quantum computations of black-hole radiation). As examples, they considered the Schwarzschild--(anti-)de Sitter and Reissner--Nordstr{\"{o}}m solutions.

We will be considering static spacetimes  and proceed purely geometrically in analogy with the uniqueness of flat foliations. As we proceed, it will be clear how this precludes the non-static case which we will discuss further in the Conclusion. We will proceed in two steps. First, to bring them on a par with flat slices, we will show that constant-curvature hypersurfaces are unique in static spherically symmetric spacetimes, from which, as in the flat hypersurface case, the new coordinates, as well as the algebraic conditions of existence, will follow (the latter of course can be identified with those obtained in \cite{Soo}). We will then show that this proof of uniqueness can be extended to prove the uniqueness of spherically symmetric spacelike hypersurfaces of arbitrary curvature in these spacetimes. This leads to the most general Painlev\`{e}--Gullstrand coordinates in static spherically symmetric spacetimes with (identical) hypersurfaces of arbitrary curvature. The original Painlev\`{e}--Gullstrand coordinates with flat slicing as well as the generalizations for constant-curvatures can be recovered as simple special cases. In all of these, as we will see, a key ingredient will be an appropriate gauge for the hypersurface metric that naturally appears in the uniqueness proof.
\\
\\
{\bf{Theorem 3.1}}: Constant-curvature spacelike hypersurfaces in spherically symmetric static spacetimes are unique up to time translation and rotation.
\\
\\
{\bf{Proof}}: We recall some standard facts about constant curvature and symmetry. As is well known, the condition of constant curvature,
\begin{equation}
{R}_{\mu\nu\rho\sigma}=\kappa ({g}_{\mu\sigma}{g}_{\nu\rho}-{g}_{\mu\rho}{g}_{\nu\sigma}), \label{curvcond}
\end{equation}
where $\kappa$ is a constant, is equivalent to maximal symmetry. For $\kappa= 0$ , $\kappa >0 $, and $\kappa <0$, these spaces are, respectively, locally isometric to the maximally symmetric spaces $\mathbb{S}^{d}$, $\mathbb{R}^{d}$, and $\mathbb{H}^{d}$, endowed with standard line elements. In addition, the maximal symmetry means all these spaces admit codimension-one spherical symmetry, which allows us to write a cohomogeneity-one metric ansatz for them\footnote{Although equivalent, as we will see, this gauge will be more beneficial than the ``proper distance
gauge"
\begin{equation}
ds^2= d\rho^2+R(\rho)^2 d\Omega_{d-1}^2,\label{hypersur2}
\end{equation}
used by other authors.}:
\begin{equation}
ds^2=a(r)^2dr^2+r^2 d\Omega_{d-1}^2.\label{hypersur1}
\end{equation}
The constant-curvature condition (\ref{curvcond}) then uniquely returns\footnote{This form of the metric is often used in Friedmann--Lema\^{i}tre--Robertson--Walker cosmological models.}
\begin{equation}
ds^2=\frac{1}{1-\kappa r^2} dr^2+r^2 d\Omega_{d-1}^2.\label{hypersurface}
\end{equation}
For $\kappa=0$, it just the flat metric in polar coordinates on $\mathbb{R}^{d}$ and, for $\kappa=1$ and $\kappa=-1$, for example, it is isometric to the familiar standard metrics on $\mathbb{S}^{d}$ and $\mathbb{H}^{d}$ with $r=\sin\chi$ and $r=\sinh\chi$.

The static spherically symmetric spacetime (\ref{GenStat}) has codimension-two spherical symmetry. An arbitrary cohomogeneity-one spherically symmetric hypersurface in it should relate the $t$ and $r$ coordinates:
\begin{equation}
t=F(r). \label{hypersurface1}
\end{equation}
This describes a curve in the $t$--$r$ plane, at each point of which a sphere of radius $r$ is attached. The induced metric on the hypersurface,
\begin{equation}
ds^2_{d}=\left(\frac{1}{g(r)}- f(r) F'^2 (r)\right) dr^2 + r^2 d\Omega_{d-1}^2, \label{hypersurface4}
\end{equation}
will be of constant curvature if (\ref{hypersurface1}) equates to (\ref{hypersurface}), i.e., if the coordinates on the spheres are identified and if
\begin{equation}
F'^2{(r)}=\frac{1}{f}\left(\frac{1}{g}-\frac{1}{1-\kappa r^2}\right).\label{thehypequation}
\end{equation}
This expression, when non-negative, can, in principle, be integrated to find $F(r)$ for either sign of the square root. Fixing the sign and defining the constant of integration to be the new time coordinate:
\begin{equation}
\overline{t}= t+\int \sqrt{\frac{1}{f}\left(\frac{1}{g}-\frac{1}{1-\kappa r^2}\right)} dr,
\end{equation}
renders (\ref{GenStat}) to
\begin{equation}
ds^2= -f(r) d{\overline{t}}^2+ 2\sqrt{f\left(\frac{1}{g}-\frac{1}{1-\kappa r^2}\right)} d{\overline{t}} dr+ \frac{1}{1-\kappa r^2} dr^2+r^2 d\Omega_{d-1}^2.\label{PGConstcur0}
\end{equation}
Thus, each constant-curvature hypersurface (\ref{hypersurface1}) in (\ref{GenStat}) corresponds to a different value of $\overline{t}$ in (\ref{PGConstcur0}) and \emph{vice versa}, thus proving uniqueness up to translation in $\overline{t}$ (or $t$) and rotation. $\Box$
\\
\\
{\it{Remark 3.1}} The above proof shows that one can dispense with the additional spherical-symmetry assumption for the constant-curvature hypersurfaces, including in the zero-curvature case (see \cite{Beig-Siddiqui}).
\subsection{Existence and Foliation}
The $\overline{t} =\mathrm{const.}$ hypersurfaces are unique. However, they can be unique only when they exist, for which one requires
\begin{equation}
\frac{1}{f}\left(\frac{1}{g}-\frac{1}{1-\kappa r^2}\right) \ge 0. \label{constcurfoli}
\end{equation}
This can be satisfied in more than one way. For $f(r)>0$ and $f(r)<0$, this requires $g(r)<1-\kappa r^2$ and $g(r)>1-\kappa r^2$, respectively, which will, in general, place restrictions on the admissible values of $r$. \emph{Once (\ref{constcurfoli}) is satisfied for all possible values of $r$ as in the original static spacetime metric (\ref{GenStat}), one would have a foliation by constant curvature hypersurfaces.}
\subsection{A Simple Example}
For the Minkowski space metric in polar coordinates
\begin{equation}
ds^2= -dt^2+dr^2+r^2 d\Omega_{d-1}^2, \label{minkpol}
\end{equation}
$g=1$. Thus, for flat hypersurfaces $\overline{t}=t$, (\ref{minkpol}) is its own Painlev\`{e}--Gullstrand form. For constant-curvature hypersurfaces, one obtains
\begin{equation}
\overline{t}= t+\int \sqrt{\frac{-\kappa r^2}{1-\kappa r^2}} dr,
\end{equation}
and
\begin{equation}
ds^2= -d{\overline{t}}^2+ 2 \sqrt{\frac{-\kappa r^2}{1-\kappa r^2}} d{\overline{t}} dr+ \frac{1}{1-\kappa r^2} dr^2+r^2 d\Omega_{d-1}^2.\label{PGConstcur}
\end{equation}
This clearly shows that $\kappa >0$ is not permissible referring to the well-known impossibility of having a $d$-sphere in $(d+1)$-dimensional Minkowski spacetime. However, $\kappa < 0$ is certainly possible as is $\kappa= 0$.
\section{Spherically Symmetric Hypersurfaces of Arbitrary Curvature}
We now turn to spherically symmetric hypersurfaces of arbitrary curvature. This means we consider hypersurfaces with metric
\begin{equation}
ds^2=a^2(r) dr^2+r^2 d\Omega_{d-1}^2,\label{hypersur1arb}
\end{equation}
where $a(r)$ is arbitrary. A
\begin{equation}
t=F(r) \label{hypersurgen}
\end{equation}
curve in (\ref{GenStat}) would give (\ref{hypersur1arb}), if, analogous to (\ref{thehypequation}),
\begin{equation}
F'^2{(r)}=\frac{1}{f}\left(\frac{1}{g}- a^2 \right).\label{thehypequationarb}
\end{equation}
The corresponding time coordinate is then
\begin{equation}
\overline{t}= t+\int \sqrt{\frac{1-a^2 g}{f g}} dr \label{newtimearbitrary}
\end{equation}
with the generalized Painlev\`{e}--Gullstrand
\begin{equation}
ds^2= -f(r) d{\overline{t}}^2+ 2\sqrt{\frac{f(1-a^2 g)}{g}} d{\overline{t}} dr+ a^2(r) dr^2+r^2 d\Omega_{d-1}^2.\label{mostgeneral}
\end{equation}
From this, the existence condition follows:
\begin{equation}
\frac{f(1-a^2 g)}{g} \ge 0.\label{finalexistence}
\end{equation}
If (\ref{finalexistence}) holds for all possible values of $r$ in the original spacetime metric (\ref{GenStat}), one has a foliation of the latter in terms of non-constant-curvature hypersurfaces (\ref{hypersur1arb}). Thus, (\ref{mostgeneral}) is the most generalized form of the Painlev\`{e}--Gullstrand coordinates for static spherically symmetric spacetimes. Note that $\int a(r)\, dr= \rho$ and $r=R(\rho)$ connects our choice of gauge in {(\ref{hypersur1})} with the proper distance gauge in {(\ref{hypersur2})}, and with this one can show the
equivalence of our generalized Painlev\`e--Gullstrand metric {(\ref{mostgeneral})} with the one obtained in \cite{Soo}.
\\
\\
{\bf{Theorem 4.1}}: Up to time translation and rotation, a static spherically symmetric spacetime (\ref{GenStat}) can be uniquely foliated by (identical) spherical spacelike three-geometries (\ref{hypersur1arb}) provided (\ref{finalexistence}) holds.
\\
\\
The remarkable similarity of (\ref{mostgeneral}) with the original Painlev\`{e}--Gullstrand coordinates (the $a=1$ case of the former) stems from our being able to use the same radial coordinate for both the hypersurface and the spacetime metrics. This also makes it is possible to obtain a number of general statements and we will note a few of them here. If a spacetime admits a flat foliation (i.e., $g \le 1$), then it would also admit a non-flat foliation provided one takes $a^2 \le 1$ everywhere on the hypersurface. This is certainly the case for constant negative curvature in which $a^2={1}/{(1+r^2)}$. The independent components of the Ricci tensor are: $R_{11}={2a'}/{(r\,a)}$ and $R_{22}= {(a'r+a^3-a)}/{a^3}$, where the derivative is with respect to $r$. Thus, it is consistent to take $a' \le 0$ and $a^2 \le 1$. With these two bounds on $a$ and $a'$ anywhere on the spacelike geometry, the eigenvalues of the Ricci curvature tensor are non-positive everywhere. Thus, we have the following: \\
\\
{\bf{Theorem 4.2}}: If a spherically symmetric static spacetime admits a flat foliation, it necessarily admits foliations by hypersurfaces whose Ricci curvature tensor is bounded above by zero.
\subsection*{Special Case: $a^2=\mathrm{const.}$}
The generalization corresponding to $a^2=p$ in (\ref{mostgeneral}), where $p$ is a constant, was obtained by others. In particular, \cite{Martel:2000rn} obtained it for a general static metric (their Eq. (4.3)) as well as for the Schwarzschild metric (their Eq. (3.6)) in which case one is limited to $0 \le p \le 1$, as can be easily verified from above. See also \cite{Lake:1994zq}, and other references in \cite{Martel:2000rn}. Note that $a=p=0$ gives the Eddington--Finkelstein coordinates.
\section{Embedding of Spacelike Spherical Geometries}
For a \emph{single}, arbitrary spherical spacelike geometry (\ref{hypersur1arb}), Theorem 4.1 can be read as a theorem on embedding. It tells us that a spherical spacelike geometry is embeddable in a one-dimensional higher static spherically symmetric spacetime (up to time translation and rotational symmetry, i.e., it is a nonrigid embedding) provided condition (\ref{finalexistence}) holds. However, there is a notable difference: now the existence condition (\ref{finalexistence}) is required to hold \emph{on the spherical spacelike geometry} (and not on the entire spacetime as for foliation).

The embedding of spherical spacelike three-geometry in Schwarzschild spacetime has been studied in the Hamiltonian formulation in \cite{OMurchadha:2003nwp}. We will now see how the same results can be obtained and extended directly using our formulation. For the Schwarzschild metric (\ref{Schw}), condition (\ref{finalexistence}) means that one requires
\begin{equation}
1-a^2\left(1-\frac{2m}{r}\right) \ge 0
\end{equation}
everywhere on the spherical spacelike three-geometry for it to be embeddable in the Schwarzschild solution (\ref{Schw}). This implies that the Schwarzschild mass $m$ should satisfy
\begin{equation}
m \ge \max \left[\frac{r}{2}\left(1-\frac{1}{a^2}\right)\right].\label{masscondition}
\end{equation}
This, as can easily be checked using $\int a(r)\, dr= \rho $\vspace{1.4pt} and $r=R(\rho)$,  is precisely the condition derived in \cite{OMurchadha:2003nwp} in
the proper distance gauge:
\begin{equation}
m \ge \max \left[ \frac{R}{2}\left( 1-{R'^{2}}\right) \right].
\label{massconditionproposerdistance}
\end{equation}

The question of when a spherical spacelike slice of a Schwarzschild solution of mass $m_1$
\begin{equation}
ds^2= \frac{dr^2}{1-\frac{2m_1}{r}}+ r^2\left(d\theta^2 + \sin^2{\theta} d\phi^2\right), \label{Schhypofmassm1}
\end{equation}
can be embedded in another Schwarzschild solution of mass $m_2$, becomes a straightforward substitution of $a^2={1}/{(1-{2m_1}/{r})}$ in (\ref{masscondition}), giving
\begin{equation}
m_2 \ge m_1.\label{masscondition1}
\end{equation}
In fact, similar simplifications ensue when one considers the embedding of constant-time slices of any static spherically symmetric spacetimes into other static spherically symmetric spacetimes. For example, it is easy to check that one obtains the same condition, (\ref{masscondition1}), for embedding the constant-time slice of the Schwarzschild--(anti-)de Sitter spacetime of mass $m_1$ in Schwarzschild--(anti-)de Sitter of mass $m_2$. It is also easy to check that (\ref{masscondition1}) is a sufficient condition for embedding the constant-time slice of Schwarzschild--de Sitter of mass $m_1$ in the vacuum Schwarzschild of mass $m_2$. One can, likewise, consider other combinations of cosmological constants and masses.
\section{Conclusion}
We have seen, from pure geometric considerations, that there are endless possibilities for foliating a static spherically symmetric spacetime by (identical copies of) spherical spacelike geometries. In each case, the foliation is unique up to time translation and rotational symmetry, and one obtains a corresponding generalized Painlev\`{e}--Gullstrand coordinate system in which the spherical spacelike geometries are constant-time slices. For foliation by flat hypersurfaces, the (original) Painlev\`{e}--Gullstrand coordinates are well known, but uniqueness was established fairly recently \cite{Beig-Siddiqui}. In this paper, we started with the question of uniqueness, which leads us to the new time coordinate, (\ref{newtimearbitrary}), and, equivalently, the generalized Painlev\`{e}--Gullstrand coordinate system, (\ref{mostgeneral}). It is instructive to compare and contrast our generalization with the generalization obtained via a physical Lorentz boost in \cite{Soo}. They differ only in the gauge for the spacelike hypersurface metric, which can be traced back to how they were obtained from mathematical and physical considerations, respectively. Can our proof of uniqueness be extended to hypersurfaces in the time dependent case, i.e., when $f$ and $g$ are functions of $t$ and $r$? Starting with a presumed relationship between $t$ and $r$, like (\ref{hypersurgen}), it is easy to see that a simple integration of $F(r)$ then would not be possible since $f(r,t)$ and $g(r,t)$ would now implicitly depend on $F(r)$ itself. This will generally make it impossible to obtain an explicit $\overline{t}$ in terms of $t$ and $r$.

The (algebraic) condition for the existence of foliation can be reinterpreted for the embeddability of a single arbitrary spacelike spherical geometry. This reproduces known results for a Schwarzschild spacetime previously obtained through the study of the Hamiltonian and momentum constraints \cite{Guven:1999hc}. This is even more useful when considering static spacetimes that are not vacuum solutions of the Einstein equations, as we noted in Section 5. A special feature in our work has been the particular gauge for the spacelike geometry which naturally comes from our proof of uniqueness. It can be transformed to the proper distance gauge as needed. However, it has the advantage that it keeps the radial coordinate intact and easily reproduced special cases obtained earlier. It also streamlined things for embedding and led to some immediate conclusions.

\end{document}